\documentclass{elsart}

\usepackage{amsfonts}
\usepackage{amssymb}
\usepackage[english]{babel}
\usepackage{exscale}
\usepackage{graphicx}
\usepackage[latin9] {inputenc}
\usepackage{mathpple}

\begin{document}

\begin{frontmatter}

\title{A modified SPH approach for fluids with large density differences}

\author{Frank Ott}
and
\author{Erik Schnetter}
\ead{schnetter@uni-tuebingen.de}

\address{Institut für Astronomie und Astrophysik, Auf der
Morgenstelle, Universität Tübingen, D-72076 Tübingen, Germany}

\begin{abstract}
We introduce a modified SPH approach that is based on discretising the
particle density instead of the mass density.  This approach makes it
possible to use SPH particles with very different masses to simulate
multi-phase flows with large differences in mass density between the
phases.  We test our formulation with a simple advection problem, with
sound waves encountering a density discontinuity, and with shock tubes
containing a contact discontinuity between air and Diesel oil.  For
all examined problems where particles have different masses, the new
formulation yields better results than standard SPH.  This is also the
case for problems in which different spatial resolutions are needed
while the mass density does not change.
\end{abstract}

\begin{keyword}
SPH
\sep Smoothed Particle Hydrodynamics
\sep multi-phase flow
\sep fluid interfaces

\PACS 02.60.Cb 
\sep  02.70.Ns 
\end{keyword}

\end{frontmatter}

\section{Motivation}

SPH is a Lagrangian particle method for solving the equations of
hydrodynamics that was invented by Lucy \cite{lucy1977} and Gingold
and Monaghan \cite{gingold1977}.  Instead of discretising space with a
grid, the matter is discretised into so-called particles which move
with the fluid and do not exchange mass.  This method is especially
suited for compressible flows with irregular boundaries.  SPH has been
used for many astrophysical problems with great success, and has also
been applied to other fields of physics, such as e.g.\ the simulation
of liquids \cite{monaghan1994} and solids \cite{benz1986,benz1987}.

Due to its Lagrangian nature, simulating several non-mixing fluids is
a straightforward extension to SPH.  Each particle gets initially
marked with the phase it belongs to, and these marks do not change
with time.

We are interested in simulating compressible flows with large density
differences in contact discontinuities, e.g.\ due to an interface
between a jet of liquid near the speed of sound in a surrounding gas.
Unfortunately, standard SPH as e.g.\ presented in
\cite{monaghan1983,monaghan1992} breaks down in this case, as
explained in this text.
Previous approaches to handle large density differences in SPH 
(e.g.\ \cite{ritchie2001}) rely on ad-hoc countermeasures.


For subsonic flows, \cite{yoon1999} presents a mesh-free numerical
method for gas-liquid phase interfaces which is based on the MPS
method \cite{koshizuka1995}.  It is restricted to incompressible
flows, whereas we are interested in compressible flows.
We are also interested in similar spatial resolutions in both phases
of the flow.  This requirement is different from e.g.\ astrophysical
collapse simulations, where one usually wants a higher resolution in
the denser regions.



Another point of interest for us is having differing spatial
resolutions, i.e.\ particles with different masses, in regions where
the mass density does not change.  This has been used on several
occasions (e.g.\ in \cite{kunze2001}).  However, this can lead to
rather large errors with standard SPH.

We will in the following present a modification of SPH that interprets
certain numerical quantities in a different manner \cite{ott1999},
leading to a stable, robust, and accurate evolution even in these
cases.

\section{Describing multi-phase flows}

One commonly used way of introducing the SPH discretisation (see also
\cite{monaghan1992}) starts out by considering an arbitrary field
$f(\mathbf{x})$.  This field is first \emph{smoothed} by folding it
with a kernel $W(\mathbf{x})$, which leads to the smoothed field
$\langle f(\mathbf{x}) \rangle$
\begin{eqnarray}
   \label{eqn:folding} \langle f(\mathbf{x}) \rangle & := & \int
   d^3x'\; f(\mathbf{x}')\; W(\mathbf{x} - \mathbf{x}')
\end{eqnarray}
where the kernel $W(\mathbf{x})$ must be normalised according to $\int
d^3x\; W(\mathbf{x}) = 1$.  One usually chooses kernels that have
approximately the shape of a Gaussian, and that have compact support
for reasons of efficiency.  The size of the domain of support is
called the \emph{smoothing length} and is usually denoted with the
letter $h$.

In the next step, the smoothing integral is discretised at $N$
particle positions $\mathbf{x}_i$, which can in principle be chosen
freely, but should of course be ``reasonably'' distributed.  This
leads to the \emph{SPH approximation} $\tilde f(\mathbf{x})$
\begin{eqnarray}
   \label{eqn:sph-approx} \tilde f(\mathbf{x}) & := & \sum_{j=1}^{N}
   V_j\; f_j\; W(\mathbf{x} - \mathbf{x}_j)
\end{eqnarray}
of the field.  The volumes $V_j$ are the discrete counterparts of the
volume element $d^3x'$ in the integral above, and have to be chosen so
as to be consistent with the spatial distribution of the particles.
It is customary to assign a certain mass $m_i$ to each particle, and
then replace the volumes through
\begin{eqnarray}
   \label{eqn:sph-volume} V_i & = & \frac{m_i}{\rho_i}
\end{eqnarray}
where $\rho_i$ is the discretised mass density assigned to the
particle.  This is motivated by the fact that the particles do not
exchange mass, which leads to the time evolution equation
\begin{eqnarray}
   \label{eqn:mass-rhs} \frac{d}{dt}\; m_i & = & 0
\end{eqnarray}
making $m_i$ a natural choice for one of the primary variables.  In
order to make SPH Lagrangian, the particles have to move with the
flow, leading to
\begin{eqnarray}
   \label{eqn:pos-rhs} \frac{d}{dt}\; \mathbf{x}_i & = & \mathbf{v}_i
\end{eqnarray}
as the time evolution equation for the particle positions
$\mathbf{x}_i$.  Here $\mathbf{v}_i$ is the discretised fluid velocity
field.

Assuming that the desired spatial resolution is about the same in all
fluid phases, one would choose similar particle spacings there,
leading to similar particle volumes $V_i$.  If the mass densities in
the different fluids are about the same, no further problems arise,
and SPH as usual can be used to describe them.  However, if the
difference in mass density is large (say, about one order of magnitude
or more), then the particle masses will differ by the same factor,
leading to problems at the phase interfaces.  These problems are
mostly caused by inaccuracies in the mass density.  These inaccuracies
are substantial for largely different particle masses.  Below we
describe how they come about, and how they can be avoided.

\subsection{Standard SPH}

There exist in principle two different methods for obtaining the
discretised mass density $\rho_i$ in the ``standard'' SPH formalism.
Both start out by considering the approximate mass density $\tilde
\rho$ at the particle position $\mathbf{x}_i$.  This quantity is
obtained from eqn.\ (\ref{eqn:sph-approx}) by choosing $f(\mathbf{x})
= \rho(\mathbf{x}_i)$ and applying eqn.\ (\ref{eqn:sph-volume}),
leading to
\begin{eqnarray}
   \label{eqn:rho-approx} \tilde \rho(\mathbf{x}_i) & = & \sum_j m_j
   W_{ij}
\end{eqnarray}
where the abbreviation $W_{ij} := W(\mathbf{x}_i - \mathbf{x}_j)$ has
been introduced.

\begin{figure}
\includegraphics[width=0.5\textwidth]{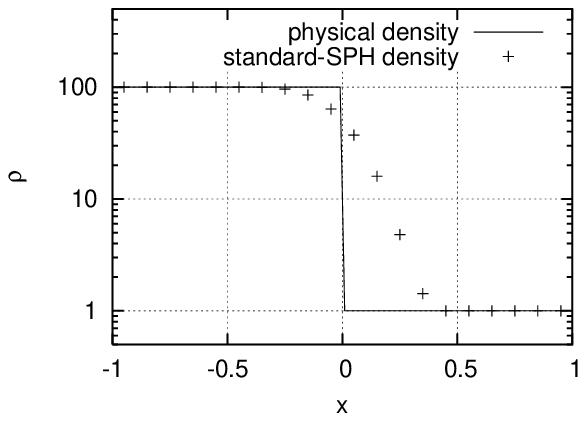}
\includegraphics[width=0.5\textwidth]{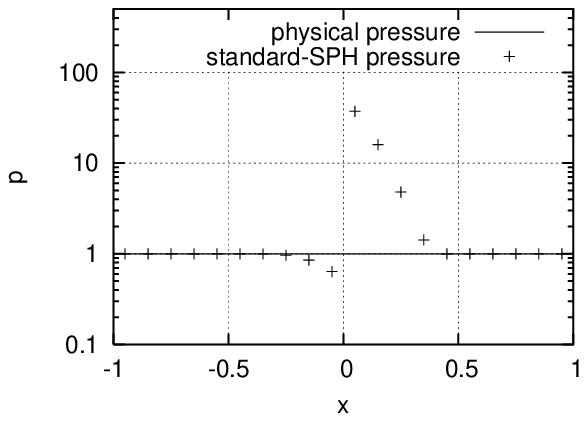}
\caption{Physical density and pressure and the corresponding
standard SPH quantities at a phase interface with equal pressure and a
density ratio of $100:1$.  Near the interface, the approximation
errors reach a factor of about 2 in the dense region and about 50 in
the thin region.  Essentially the same picture results when there is
no change in mass density, but when instead the particle mass and
spacing are changed by a factor of 100 to achieve a higher spatial
resolution (not shown).  (These approximation errors do not exist when
a different SPH formulation is used; see section
\ref{sec:modified-sph} below).}
\label{fig:wrong-pressure}
\end{figure}

The first and conceptually simpler method to obtain $\rho_i$ from this
is by setting $\rho_i := \tilde \rho(\mathbf{x}_i)$, leading to
\begin{eqnarray}
   \label{eqn:rho-direct} \rho_i & = & \sum_j m_j W_{ij}
   \quad\textrm{.}
\end{eqnarray}
This method is often used for astrophysical problems when there are
free boundaries, i.e.\ when the matter distribution extends into
vacuum.  However, it is not suited for a phase interface with a large
density discontinuity.  The smoothing inherent in eqn.\
(\ref{eqn:rho-direct}) will smooth out the density jump over a region
of the size of the smoothing length $h$ in either direction of the
discontinuity.  Particles in this region will then ``see'' a density
that is much less or much larger from the real density in their phase.
When this density is used to calculate the pressure through the two
phases' equations of state, the pressure will be very wrong (as shown
in figure \ref{fig:wrong-pressure} for two ideal gases), and it is
basically impossible to set up a stable interface in equilibrium.
This problem becomes even more severe when one of the fluids has a
stiff equation of state, as is the case e.g.\ in liquids or solids,
because then density inaccuracies will lead to even larger errors in
the pressure.

The second method for obtaining the mass density $\rho_i$ is by
integrating $\rho_i$ in time via the time derivative of eqn.\
(\ref{eqn:rho-direct}), leading to
\begin{eqnarray}
   \label{eqn:rho-rhs-monaghan} \frac{d}{dt}\; \rho_i & = & \sum_j
   m_j\; (\mathbf{v}_i - \mathbf{v}_j) \cdot \nabla W_{ij}
\end{eqnarray}
where eqn.\ (\ref{eqn:pos-rhs}) has been used, and the abbreviation
$\nabla W_{ij} := (\nabla W)(\mathbf{x}_i - \mathbf{x}_j)$ has been
introduced.  This method has the advantage that the initial data for
$\rho_i$ can be chosen freely, so that density discontinuities can be
modelled.  This can be used to simulate surfaces of liquids and
solids.

The problem with this method is similar to the problem encountered
when calculating the density directly from the particle distribution
by eqn.\ (\ref{eqn:rho-direct}).  The particles on each side of a
phase interface ``see'', via the term $m_j$ in eqn.\
(\ref{eqn:rho-rhs-monaghan}), very different particle masses on the
other side of the interface.  The values of $d/dt\; \rho_i$ then
contain large inaccuracies, leading to instabilities at phase
interfaces.

\subsection{Modified SPH}
\label{sec:modified-sph}

However, eqn.\ (\ref{eqn:rho-direct}) is not cast in stone.  An ansatz
equivalent to but different from the one leading to this equation is
not to smooth the mass density $\rho_i$, but rather the particle
density $n_i = 1/V_i$ \cite{ott1999}.  This is easily motivated by the
fact that the mass density can be discontinuous over a phase
interface, while the particle density is not, according to our
assumption of similar spatial resolutions on both sides.  Smoothing
the particle density $n_i := \tilde n(\mathbf{x}_i)$ via eqn.\
(\ref{eqn:sph-approx}) leads to
\begin{eqnarray}
   \label{eqn:vol-direct} n_i & = & \sum_j W_{ij}
\end{eqnarray}
and by taking its time derivative, the equation
\begin{eqnarray}
   \label{eqn:vol-rhs-ott} \frac{d}{dt}\; n_i & = & \sum_j
   (\mathbf{v}_i - \mathbf{v}_j) \cdot \nabla W_{ij}
\end{eqnarray}
is obtained after using eqn.\ (\ref{eqn:pos-rhs}).  As it is customary
in SPH to use $\rho_i$ instead of $n_i$, we apply eqns.\
(\ref{eqn:sph-volume}) and (\ref{eqn:mass-rhs}) and arrive at
\begin{eqnarray}
   \label{eqn:rho-rhs-ott} \frac{d}{dt}\; \rho_i & = & m_i \sum_j
   (\mathbf{v}_i - \mathbf{v}_j) \cdot \nabla W_{ij} \quad\textrm{.}
\end{eqnarray}
This new formulation of the equation of continuity is the key element
of our SPH approach.
It should be noted that this equation is identical to eqn.\
(\ref{eqn:rho-rhs-monaghan}) when all particle masses $m_i$ are the
same, which is the case for many single-phase SPH simulations.

For the simulations presented in this text, we discretise the Euler
and the internal energy equations in established ways:
\begin{eqnarray}
   \label{eqn:vel-rhs} \frac{d}{dt}\; \mathbf{v}_i & = & -
   \frac{1}{\rho_i}\; \sum_j \frac{m_j}{\rho_j}\; \left( p_j + p_i
   \right)\; \nabla W_{ij}
\\
   \label{eqn:eint-rhs} \frac{d}{dt}\; e_i & = & \frac{1}{2}\;
   \frac{1}{\rho_i} \sum_j \frac{m_j}{\rho_j}\; \left( p_j + p_i
   \right)\; (\mathbf{v}_i - \mathbf{v}_j) \cdot \nabla W_{ij}
\end{eqnarray}
where $e_i$ is the specific internal energy.  The symmetrisations
$(p_j + p_i)$ are e.g.\ explained in \cite{monaghan1992}.

\section{Tests}

In the following we test the new SPH formulation and compare it to
analytic solutions as well as simulations performed using standard SPH
as described in \cite{monaghan1992}.  That is, the only difference
between these two formulations is that we use eqn.\
(\ref{eqn:rho-rhs-ott}) instead of eqn.\ (\ref{eqn:rho-rhs-monaghan}).
This also means that other parts of an SPH code such as time
integration or artificial viscosity are not affected.  As test cases,
we use an advection problem, a sound wave encountering a discontinuous
change in the sound speed, and a shock tube with a Diesel--air
interface.

\subsection{Advection equation}

We compare the standard and the new SPH formulation by simulating a
one-dimentional advection equation.  That is, we solve the equation of
continuity for the density $\rho$ while prescribing the velocity field
$v$.  The velocity profile (which is constant in time) and the initial
density profile are given by
\begin{eqnarray}
   v(x) & = & \frac{x}{1 + q x^2}
\\
   \rho_0(x) & = & A\; x^2\; \exp \left\{ - \left( \frac{x-x_0}{W}
   \right)^2 \right\}
\end{eqnarray}
with $A=1.5$, $x_0=1$, $W=0.4$, and $q=0.2$.  We initially place the
particles with equidistant spacings with a density of $n=10$ particles
per unit length and use a smoothing length of $h=0.25$.  The particle
masses are chosen according to the initial density profile at the
initial particle positions, i.e.\ they differ.  Advection problems are
particularly well suited test problems for Lagrangian methods, so we
expect a high accuracy from this low resolution.

\begin{figure}
\includegraphics[width=0.5\textwidth]{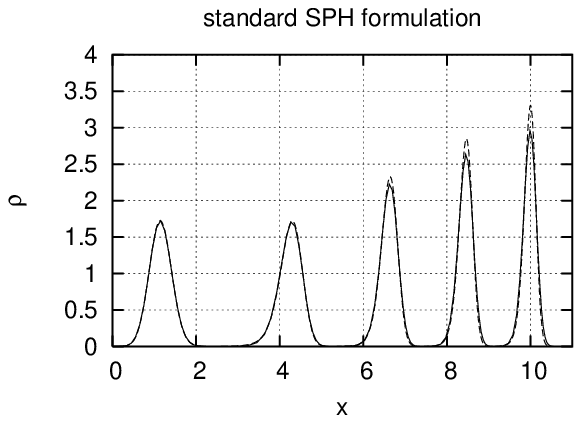}
\includegraphics[width=0.5\textwidth]{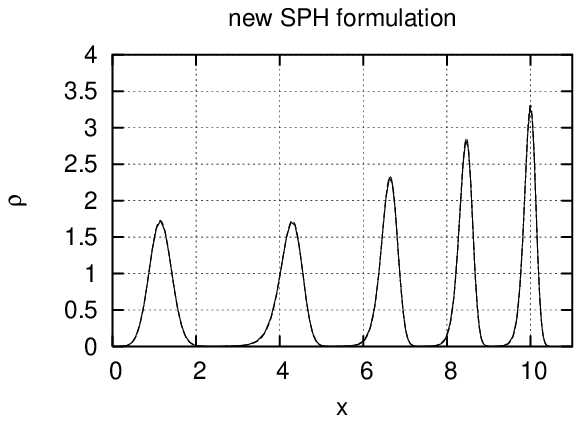}
\caption{Comparison of simulations with the standard SPH and the new
SPH formulation at five different times.  The dotted lines show the
analytic solution, the solid lines show the particle values.  Standard
SPH is less accurate at the narrower peaks, which correspond to later
times.}
\label{fig:advection-comp}
\end{figure}

The results of simulating this equation with both the standard and the
new SPH formulation are presented in figure \ref{fig:advection-comp},
which shows the density $\rho$ at five different times.  Both
formulations track the analytic solution very nicely in spite of the
coarse resolution.  However, at later times, the standard SPH
formulation underestimates the density near the peaks, while the new
formulation stays much closer to the analytic solution.

\subsection{Sound wave}

\begin{figure}
\includegraphics[width=0.5\textwidth]{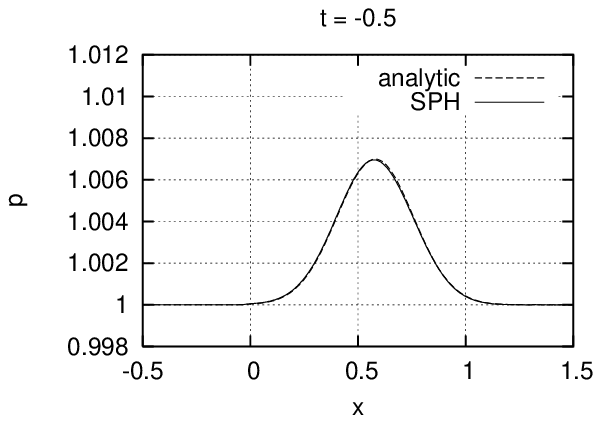}
\includegraphics[width=0.5\textwidth]{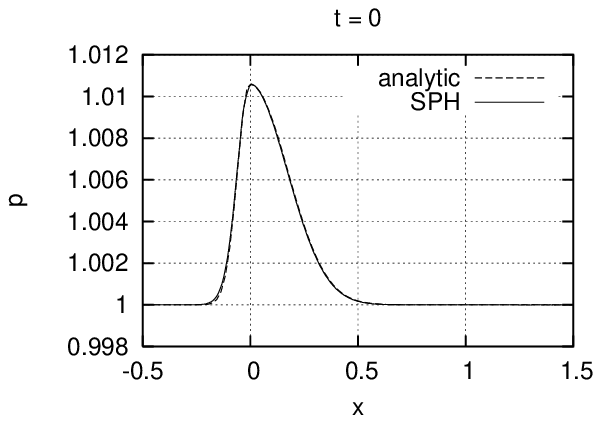}
\includegraphics[width=0.5\textwidth]{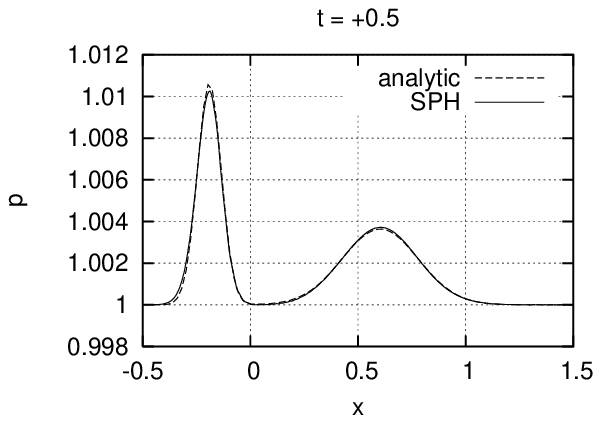}
\includegraphics[width=0.5\textwidth]{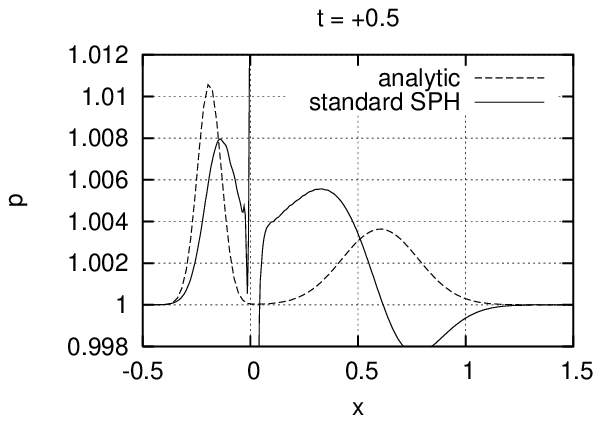}
\caption{A sound wave crossing an interface between two ideal gases
with a density ratio of $10:1$.  The interface is at $x=0$.  Shown is
the pressure at the times $t=-0.5$, $t=0$, and $t=+0.5$, where dotted
lines mark the analytic solution.  The graph in the lower right hand
corner shows the result at $t=+0.5$ of a simulation with standard SPH
for comparison.}
\label{fig:soundwave}
\end{figure}

The sound wave test case consists of two regions containing the same
ideal gas, but with different densities and in pressure equilibrium.
The density discontinuity is located at $x=0$ with a density ratio of
$10:1$.  These conditions also lead to different temperatures, and to
sound speeds with a ratio of $1:\sqrt{10}$.  Figure
\ref{fig:soundwave} shows an initially Gaussian-shaped sound wave at
different times.

At $t=-0.5$ the initial wave travels to the left.  At $t=0$ the wave
has reached the interface where it is partially transmitted and
partially reflected.  At $t=+0.5$ the wave consists of two packets,
travelling in different directions with different speeds.  The
simulation was performed with $n=200$ particles per unit length and a
smoothing length of $h=0.1$.  The analytic solution is shown as dotted
line underneath the simulation result.  The SPH simulation with the
new equation of continuity (\ref{eqn:rho-rhs-ott}) tracks the analytic
solution quite well.  On the other hand, standard SPH using eqn.\
(\ref{eqn:rho-rhs-monaghan}) performs rather poorly in this case, as
can be seen in the graph in the lower right hand corner: the
transmission and reflection coefficients are wrong, and the pressure
develops spikes at the interface.  We assume that the reason for this
is just the one demonstrated in figure \ref{fig:wrong-pressure}.

\subsection{Shock tube}

\begin{figure}
\includegraphics{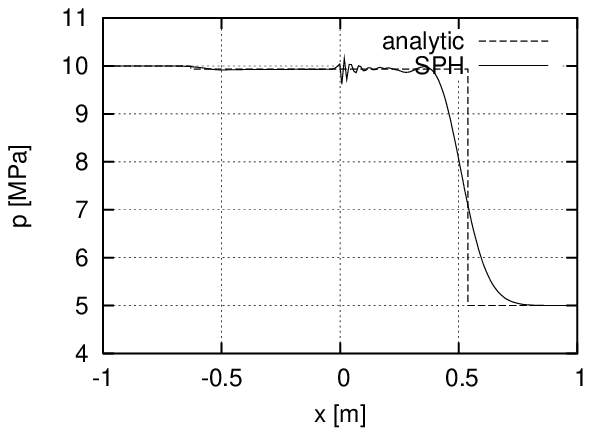}
\includegraphics{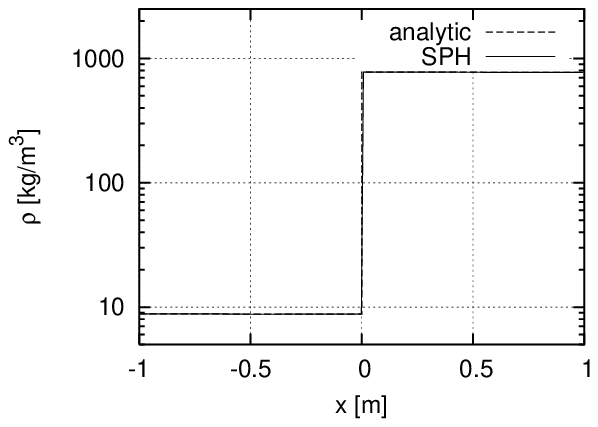}
\caption{A shock wave emanating from an interface with air to the left
and Diesel to the right.  Shown are pressure and density at $t=5
\times 10^{-4}\,\mathrm{s}$, where dotted lines mark the analytic
solution.  The discontinuity is initially at $x=0$.  The shock front
is spread out over $8$ smoothing lengths by artificial viscosity.  The
pressure difference across the rarefaction wave is rather small.  The
shock and rarefaction waves are not visible in the density graph
because of the large scale differences.  The contact discontinuity is
stable and well preserved in spite of the coarse resolution; the
pressure oscillations do not grow with time.}
\label{fig:shockwave}
\end{figure}

A further test case for our formulation is a shock tube containing a
Diesel--air interface, shown in figure \ref{fig:shockwave}.  The
initial discontinuity is at $x=0$, with air to the left and (liquid)
Diesel oil to the right.  The initial pressure ratio is $2:1$, the
density ratio about $1:80$.  Here we have artificially decreased the
air density and increased the air temperature by a factor of $10$ to
create a more difficult test case.  Table \ref{tab:initial} lists the
exact initial data for this test case.  The shock wave in the Diesel
oil travels to the right, the rarefaction wave in the air to the left.
Because the Diesel oil is nearly incompressible, the final pressure is
close to the initial pressure in the air phase.  The equation of state
for the Diesel oil was kindly provided to us by the Robert Bosch GmbH.

\begin{table}
\caption{Initial data for the shock tube test case}
\label{tab:initial}
\begin{tabular}{rl|ll}
\multicolumn{2}{c|}{Quantity} & air & Diesel oil \\\hline
$\rho$ & $[\mathrm{kg/m^3}]$ & $8.81$ & $772.546$ \\
$p$ & $[\mathrm{MPa}]$ & $10$ & $5$ \\
$T$ & $[\mathrm{K}]$ & $3931.5$ & $393.15$
\end{tabular}
\end{table}

The simulation was performed with $n=100$ particles per metre
and a smoothing length of $h=5 \times 10^{-2}\,\mathrm{m}$.  We use
the artificial viscosity presented in \cite{monaghan1992} with a
viscosity coefficient of $\alpha=0.5$, because some artificial
viscosity is necessary to produce entropy in the shock front.  This
spreads out the shock front over about 8 smoothing lengths, which is
acceptable for our purposes.

The initial pressure discontinuity remains visible as spikes at the
contact discontinuity.  These spikes are caused by the numerical
initial data, which have a discontinuity and hence contain high
frequency modes that are not resolved in the simulation.  According to
eqn.\ (\ref{eqn:folding}), the initial data should be smoothed before
the SPH formalism is applied.  We skip this step because we want to
show that these high frequency modes do not harm the simulations.
They remain present, but are not amplified.  The formulation is
stable.  Table \ref{tab:result} compares several important quantities
of the simulation result to the analytic solution, showing very good
agreement in the shock relations.

\begin{table}
\caption{Comparison of the analytic solution and the simulation result
for the shock relations in the Diesel phase.  $v_s$ is the shock front
speed, $v_D$ the post-shock Diesel speed.  The sound speed in the
pre-shock Diesel phase is $1059.6\,\mathrm{m/s}$.}
\label{tab:result}
\begin{tabular}{rl|ll}
\multicolumn{2}{c|}{Quantity} & analytic & SPH
\\\hline
$v_s$ & $[\mathrm{m/s}]$ & $1077$ & $1056 \pm 20$
\\
$v_D$ & $[\mathrm{m/s}]$ & $5.93$ & $5.94 \pm 0.02$
\\
$\Delta\rho_D$ & $[\mathrm{kg/m^3}]$ & $4.27$ & $4.28 \pm 0.02$
\\
$\Delta p_D$ & $[\mathrm{MPa}]$ & $4.93$ & $4.94 \pm 0.02$
\\
$\Delta T_D$ & $[\mathrm{K}]$ & $0.860$ & $0.855 \pm 0.005$
\end{tabular}
\end{table}

We did not manage to perform this simulation with standard SPH.  As
illustrated in figure \ref{fig:wrong-pressure}, the error in the
density near the contact discontinuity leads to an error in the
pressure.  The pressure error can be orders of magnitude larger than
the pressure itself for fluids which have a stiff equation of state.
This problem does not exist in the new formulation.

\section{Conclusion}

We describe a modification to the standard SPH formalism that smoothes
the particle density instead of the mass density.  As tested by
simulating an advection equation, sound waves, and shock waves, the
new formulation either yields more accurate results than standard SPH,
or the equivalent simulation with standard SPH is not stable.  We
conclude that this modified SPH formulation is an effective method for
simulating multi-phase flows, and conjecture that this modification is
beneficial for all simulations where particles have different masses.

The authors would like to thank Prof.\ Hanns Ruder for encouraging
this work.  We would also like to thank the Robert Bosch GmbH for
providing an equation of state for Diesel oil.  Financial support was
provided by the Robert Bosch GmbH and the Ministerium für Wissenschaft
und Forschung in Baden-Württemberg.

\bibliographystyle{elsart-num}
\bibliography{twophase}

\end{document}